\documentclass[twocolumn,showpacs,preprintnumbers,amsmath,amssymb,superscriptaddress,revsymb,prb]{revtex4}

\usepackage[dvips]{graphicx}
\usepackage{mathptmx}
\usepackage{dcolumn}

\newcommand{\etal}{\emph{et al.}}

\newcolumntype{.}{D{.}{\cdot}{3.10}}

\begin{document}

\title{Lithiation of silicon via lithium Zintl-defect complexes}


\author{Andrew J. Morris\footnote{Email:
    ajm255@cam.ac.uk}\footnote{Present Address: Cambridge University
    Nanoscience Centre, 11 J.~J. Thomson Avenue, Cambridge CB3 0FF,
    United Kingdom}}

\affiliation{Department of Physics and Astronomy, University College
  London, Gower St, London WC1E 6BT, United Kingdom}

\author{R. J. Needs} 
\affiliation{Theory of Condensed Matter Group, Cavendish Laboratory,
  University of Cambridge, J. J. Thomson Avenue, Cambridge CB3 0HE,
  United Kingdom}

\author{Elodie Salager} 

\author{C. P. Grey} \affiliation{Department of Chemistry, University
  of Cambridge, Lensfield Road, Cambridge CB2 1EW, United Kingdom}

\author{Chris J. Pickard}

\affiliation{Department of Physics and Astronomy, University College
  London, Gower St, London WC1E 6BT, United Kingdom}

\date{\today{}}

\begin{abstract} 
  An extensive search for low-energy lithium
  defects in crystalline silicon using density-functional-theory methods and the
  \emph{ab initio} random structure searching (AIRSS) method shows
  that the four-lithium-atom substitutional point defect is
  exceptionally stable.  This defect consists of four lithium atoms
  with strong ionic bonds to the four under-coordinated atoms of a
  silicon vacancy defect, similar to the bonding of metal ions in
  Zintl phases.  This complex is stable over a range of silicon
  environments, indicating that it may aid amorphization of crystalline silicon and form upon delithiation of the silicon anode of a Li-ion rechargeable battery.
\end{abstract}

\pacs{61.05.-a, 61.72.J-, 82.47.Aa}
\maketitle

Monovacancies in silicon may be retained after growth with
concentrations of about $3\times10^{16}$
cm$^{-3}$. \cite{dannefaer:JAP:1995} Lithium present in the bulk is
strongly attracted to vacancies, although there is a substantial
energy barrier to forming a lithium substitutional defect in the perfect
crystal. \cite{dannefaer:JAP:1995}
In this article we denote a defect complex by listing its constituents
between braces (\{...\}).
For example, the lithium substitutional defect is denoted by
\{Li,$V$\} since it comprises a lithium atom within a silicon
vacancy.
A lithium-vacancy complex was assigned to an electron paramagnetic
resonance (EPR) center by Goldstein \cite{Goldstein:PRL:1966} and
lithium drift rate experiments have also shown that lithium binds to
vacancies, preventing further mobility. \cite{Klimkova:PSSa:1970} Our
density-functional theory (DFT) calculations show that the formation
of a Frenkel-type impurity defect (\{Li,$V$,$I$\}) from a $T_d$
interstitial Li atom requires an energy of 3.73\,eV, which is in
excellent agreement with the value of 3.75\,eV calculated by Wan
\etal\ \cite{wan:JPCM:2010}
Wan \etal\ suggested that the substitutional lithium defect,
\{Li,$V$\} will not form due to its large formation energy.
However, we calculate its formation energy from bulk silicon and
lithium metal to be 3.09\,eV, and hence \{Li,$V$\} is more stable than
a separated vacancy and lithium interstitial (\{$V$\} and \{Li\}).
It is likely that silicon vacancies play a significant role in the
interaction between lithium and silicon.

Along with the wide range of technological uses of silicon, such as
semiconductor devices and photovoltaics, it has recently been proposed
as an anode for lithium-ion batteries.
Lithium intercalated graphite is the standard lithium-ion battery
(LIB) negative electrode material, due to its good rate capability and
cyclability, but demand for even higher performance LIBs has motivated
the investigation of other materials.
Silicon is an attractive alternative since it has ten times the
gravimetric and volumetric capacity of graphite (calculated from the
initial mass and volume of silicon) but, unlike graphite, silicon
undergoes structural changes on
lithiation. \cite{lai:JES:1976,wen:JSSC:1981,weydanz:JPS:1999}

In the first stages of lithiation, in which silicon atoms greatly
outnumber lithium, it has been proposed that lithiation occurs as
interstitial lithium defects form near the silicon
surface. \cite{peng:JPC:2010}
Previous theoretical studies have shown that a single lithium atom in
bulk silicon resides at the $T_d$ site and that at higher
concentrations lithium clusters can promote the breaking of
silicon-silicon
bonds. \cite{wan:JPCM:2010,kim:JPCC:2010,watkins:PRB:1970}
There are equal numbers of $T_d$ sites and silicon atoms in $c$-Si
and, since $a$-Li$_y$Si forms at $y\approx 0.3$ in micron-sized
silicon clusters (a Li:Si ratio of $\approx1:3$
\cite{key:JACS:2009,li:JES:2007}), the presence of lithium must lead
to the breaking of silicon-silicon bonds before all $T_d$ sites are
filled with lithium.
Chan \etal\ presented a density-functional theory study of the
lithiation of silicon surfaces and $c$-Si in which they described the
charging or discharging of the half-cell by the addition or
subtraction of a succession of lithium
impurities. \cite{chan:JACS:2012}
Their study suggests a lithiation pathway from $c$-Si to
Li$_{15}$Si$_4$ and correctly predicts enhanced expansion in the (110)
direction, as observed by Goldman \etal \cite{goldman:AFM:2011}
Understanding Li defects in silicon is a vital step towards a full
appreciation of the lithiation of a Si electrode in a LIB.

Our previous study showed that interstitial lithium binds to any
hydrogen present and breaks silicon-silicon
bonds. \cite{morris:PRB:2011}
The dynamics of lithium in the bulk are beyond the scope of this
study, but understanding the underlying stable lithium configurations
is crucial to any subsequent dynamical study.
In this article we show that the \{4Li,$V$\} complex is the most
stable defect when silicon vacancies are present.
Below we describe briefly our computational search method, followed by
a description of the \{4Li,$V$\} defect and its behavior in silicon
with a range of silicon environments and lithium chemical potentials.

\begin{figure}
\includegraphics*[width=80mm]{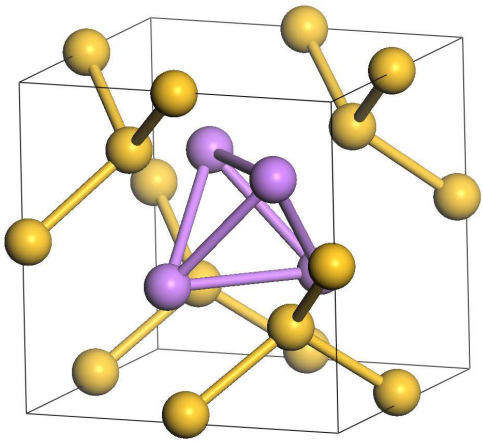}
\caption[]{(Color online) The \{4Li,$V$\} defect of $T_{d}$ symmetry
  in an 8 atom conventional cell.  The lithium atoms form a
  tetrahedron within the silicon vacancy.  Lithium atoms are in pink
  (dark grey), silicon in yellow (light grey). The bonds between the
  lithium atoms are to guide the eye.}
\label{Fig:4Li_s}
\end{figure}

\begin{figure}
\includegraphics*[width=80mm]{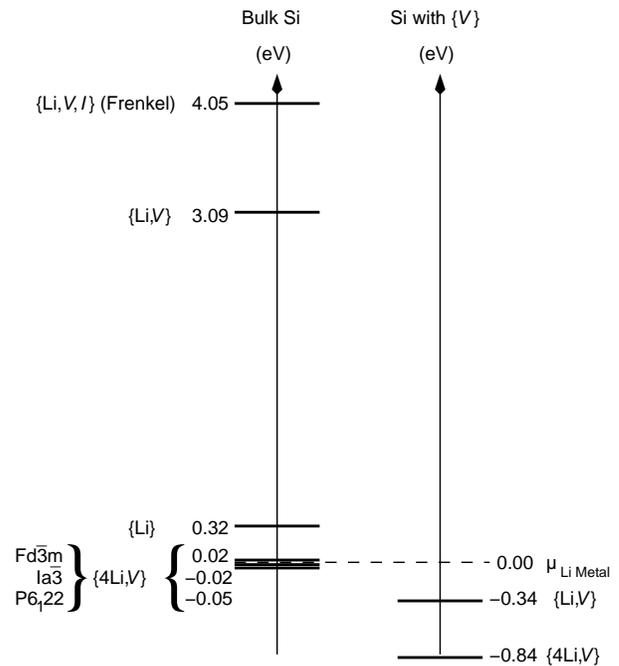}
\caption[]{The chemical potential of lithium in point defects in
  silicon relative to lithium metal, $\mu_{\mathrm{Li\ metal}}$. Any
  complex lower in energy than $\mu_{\mathrm{Li\ metal}}$ is likely to
  form in silicon from a reservoir of lithium metal and therefore may
  be found in silicon electrodes on initial lithiation.  The left-hand
  scale shows the chemical potential of lithium defects in bulk
  silicon; the right-hand scale shows the chemical potential of
  lithium defects when vacancies are also present in the silicon
  lattice.  The lithium substitutionals \{Li,$V$,$I$\} and \{Li,$V$\}
  are unlikely to form in bulk silicon due to their high energy, but
  lithium will bind to vacancies present in the bulk forming
  \{Li,$V$\} (see right-hand scale).  The most likely complex to form,
  whether or not vacancies are present, is \{4Li,$V$\}.  The
  \{4Li,$V$\} complex is almost degenerate with $\mu_{\mathrm{Li\
      metal}}$ in diamond-structure ($Fd\overline{3}m$) silicon that
  initially contains no vacancies, however, \{4Li,$V$\} is energetically
  favorable in the $P6_122$ and $Ia\overline{3}$ lower-symmetry
  tetrahedrally-bonded silicon phases.  Hence at room temperature it
  is very likely that \{4Li,$V$\} will form in a range of silicon
  environments, breaking silicon-silicon bonds and aiding the
  integration of lithium into the silicon lattice. }
\label{Fig:Energy}
\end{figure}

\emph{Ab initio} random structure searching (AIRSS)
\cite{Pickard_AIRSS} has been successful in predicting the
ground-state structures of point defects (including the interaction of
impurities with vacancies) in semiconductors
\cite{morris:PRB:2008,morris:PRB:2009} and
ceramics. \cite{mulroue:PRB:2011,mulroue:PRB:2011:He}
The initial structures for the searches were prepared as follows.
A silicon atom and its four nearest neighbors were removed from a
periodic supercell of diamond-structure ($Fd\overline{3}m$) silicon
containing 32 atomic sites; this has previously been shown to be large
enough to accommodate impurities of this size at the searching
stage. \cite{morris:PRB:2008}
Four silicon atoms, along with either 1, 2, 3, 4 or 5 lithium atoms,
were then randomly placed within a sphere of radius 4\,\AA\ centered
on an atomic site.
We relaxed approximately 400 different starting structures.
Each structure was relaxed, while keeping the supercell fixed, until
the DFT forces on the ions were smaller than 0.05\,eV/\AA.
A few searches were also carried out adding hydrogen atoms along with
lithium over a range of Li/H combinations.

The plane wave \textsc{castep} \cite{CASTEP:ZK:2004} DFT code and the
PBE (Perdew-Burke-Ernzerhof) exchange-correlation functional were used
with ``ultrasoft'' pseudopotentials and a basis set containing plane
waves with energies of up to 300 eV.
The Brillouin zone (BZ) was sampled using 2$\times$2$\times$2
$k$-points arranged according to the multiple-$k$-point generalization
of the Baldereschi scheme. \cite{morris:PRB:2008}
The candidate structures were then further relaxed to higher accuracy
in larger supercells containing 256 atomic sites and a standard
2$\times$2$\times$2 Monkhorst-Pack (MP) grid corresponding to a BZ
sampling grid finer than $2\pi\times0.033$\,\AA$^{-1}$.

We found, in order of increasing stability, \{Li,$V$\}, \{2Li,$V$\},
\{3Li,$V$\} and \{4Li,$V$\} complexes.
The \{4Li,$V$\} defect with $T_d$ symmetry, shown in
Fig.\,\ref{Fig:4Li_s}, was the lowest energy Li/Si-vacancy complex
found.
This complex was re-computed allowing spin polarization, but no spin
moment was obtained with the PBE density functional.
The four lithium atoms are at the vertices of a tetrahedron, with
edges of 2.79\,\AA\, so that the Li-Li bonds are longer and therefore
weaker than those in the lithium dimer (2.672\,\AA) but shorter than
those in bulk lithium (2.98\,\AA).
Each lithium atom has three nearest-neighbor (NN) silicon atoms at a
distance of 2.51\,\AA\ and three second NNs 2.71\AA\ away, compared to
first and second NNs for the relaxed \{Li\} defect at 2.46\,\AA\ and
2.75\,\AA, respectively. \cite{kim:JPCC:2010}
Canham \cite{Canham:Physica:1983} associated the photoluminescent
``Q''-line of lithiated silicon with a \{4Li,$V$\} defect, and
reported a symmetry lowering to $C_{3v}$, but we did not find any
\{4Li,$V$\} complexes of this symmetry in our AIRSS searches or by
breaking the $T_d$ symmetry ``by hand''.

Using the gauge-including projector-augmented wave method
(GIPAW)\cite{pickard:PRB:2001,yates:PRB:2007} implemented in
\textsc{castep} the chemical shift of lithium in the \{4Li,$V$\}
complex is 9\,ppm.
This makes the \{4Li,$V$\} complex very difficult to differentiate
from the lithium clusters that form in $a$-Li$_y$Si, which have broad
peaks in the chemical shift range +20 to -10\,ppm.\cite{key:JACS:2009}
The isotropic chemical shielding of lithium in \{4Li,$V$\} was
obtained in the 32 silicon atomic site simulation cell using a BZ
sampling finer than $2\pi\times0.03\,$\AA$^{-1}$ and a well-converged
basis set containing plane waves up to $850$\,eV.
The chemical shift was obtained by comparing to the calculated value
in crystalline LiF which has a known chemical shift of
$\delta=-1$\,ppm relative to the standard 1M LiCl solution (at
0\,ppm).
In silicon there is a significant change in lattice constant depending
on the exchange-correlation functional chosen. However using the LDA,
GGA and experimental lattice parameters changed the lithium chemical
shift of the \{4Li,$V$\} defect by less than 2\,ppm.

The lithium and silicon atoms in the \{4Li,$V$\} complex bond in a
similar way to a Zintl phase.
Zintl phases consist of an alkali metal or alkaline-earth metal and an
element from the $p$-block of the periodic table known as the
\emph{Zintl ion}.
The ions are bound electrostatically as the metal ion is completely
ionized by the more electronegative Zintl ion.
A Zintl ion with atomic number $Z_{\rm zi}$, which gains $n$ electrons
from the metal, requires the donation of $n$ fewer electrons through
covalent bonding to satisfy the ``octet rule''.
The Zintl ion behaves isoelectronically with the element of atomic
number $Z_{\rm zi}+n$.
For example, the phosphorus ($Z=15$) atoms in Li$_1$P$_1$ gain one
electron and form long 2-fold coordinated chains similar to those
found in sulfur ($Z=16$) compounds.\cite{ivanov:ACIE:2012}

\begin{figure}
\includegraphics*[width=80mm]{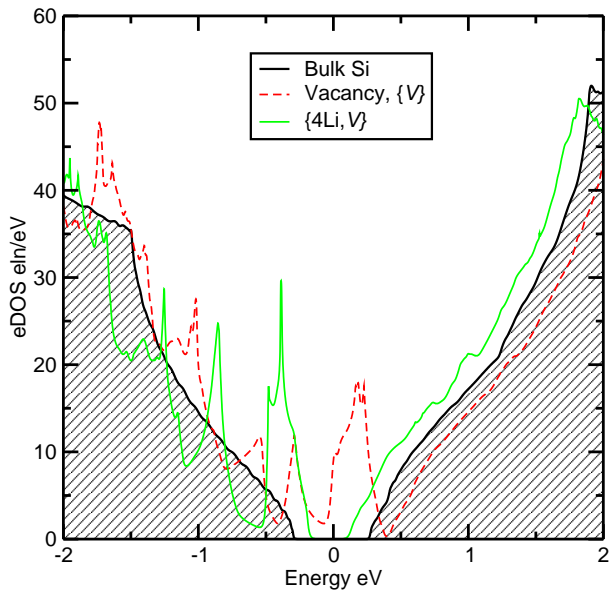}
\caption[]{(Color online) The electronic density of states of the
  \{4Li,$V$\} defect around the Fermi level (set to 0\,eV) calculated
  using \textsc{OptaDOS} (green line).  The \{$V$\} defect (red
  dotted) forms a state in the band gap which is subsequently filled
  by the electrons donated from the lithium atoms.}
\label{Fig:DOS}
\end{figure}

%
As in the lithium/phosphorus model compounds mentioned above, Mulliken
population analysis shows that the lithium atoms in \{4Li,$V$\} have a
positive charge, and the valence charge density shows a lone pair of
electrons on the four three-fold covalently bonded silicon atoms (in
this case making the four silicon atoms isoelectronic to phosphorus).
The lithium-silicon interaction is not well-described as a covalent
bond, since each lithium atom has 3 silicon NN each with a small
($0.18|e|$) charge overlap.
We believe that the fulfillment of the Zintl-rules leads to the
stability of the \{4Li,$V$\} defect.

The single silicon vacancy, \{$V$\}, has an unoccupied electronic
state in the band gap, in contrast \{4Li,$V$\} has a similar-shaped
density of electronic states (DOS) but with the gap state filled, see
Fig.\,\ref{Fig:DOS}.
Projecting the DOS onto local orbitals using \textsc{OptaDOS} reveals
that the gap states have a silicon-like character, indicating charge
transfer from lithium to silicon.\cite{optados}

A selection of charged cells (2-, 1-, 1+ and 2+) were considered in
supercells containing 8, 64, 216 and 512 silicon atomic sites.
The atomic positions were optimised after initially breaking the spin
symmetry and adding a neutralising background charge.
The charge did not localize around the defect in any of these
calculations, indicating that the charged states are unstable, and the
neutral defect is therefore the only stable state of \{4Li,$V$\}.

We now turn our attention to the energetics of \{4Li,$V$\}.
The formation energy per atom of the defect is calculated as
\begin{equation}
E_f/A=\left ( E - 4\mu_{\rm Li} - n_{\rm Si}\mu_{\rm Si} \right ) /4,
\end{equation}
where $E$ is the total energy of the defect cell, $n_{Si}$ is the
number of silicon atoms present in the supercell, and $\mu_{Li}$ and
$\mu_{Si}$ are the chemical potentials of lithium and silicon,
respectively, which are discussed in detail below.
When considering the energetics of a defect it is important to compare
its stability against physically meaningful chemical potentials.
We consider three scenarios: (I) we model silicon prior to
lithiation, and therefore both $\mu_{\mathrm{Si}}$ and
$\mu_{\mathrm{Li}}$ are derived from their bulk elemental compounds;
(II) on commencement of lithiation the silicon contains mobile lithium
atoms, therefore $\mu_{\mathrm{Si}}$ is derived from bulk Si and
$\mu_{\mathrm{Li}}$ from the \{Li\} defect at the $T_d$ site; and
(III) is as (I) but, since monovacancies may be found in bulk silicon,
$\mu_{\mathrm{Si}}$ is derived from bulk Si with an isolated vacancy
and $\mu_{\mathrm{Li}}$ is taken from Li metal.
A 4$\times$4$\times$4 MP grid was used for calculating the defect
formation energies in the 256 atom cells, which gives a $k$-point
spacing of less than $2\pi\times0.016$\,\AA$^{-1}$.
The formation energy of the $T_d$ \{4Li,$V$\} complex in scenario (I)
is 0.02\,eV per Li atom, which is a small positive formation energy,
so it seems likely that this complex will form at room temperature.
In the type (II) scenario the formation energy is -0.30\,eV per Li
atom, as shown on the left-hand scale of Fig.\,\ref{Fig:Energy}, (the
difference in formation energy between the \{Li\} and \{4Li,$V$\}
defects in diamond-structure silicon), implying that it is
energetically favorable for mobile lithium atoms present in the bulk
to ``dig out'' a silicon vacancy.

Scenario (III) requires the calculation of the structure and
energetics of the silicon vacancy.
When using a simulation cell of 256 sites or more we found that the
single silicon vacancy \{$V$\} lowers its symmetry from $T_d$ to
$D_{2d}$.\cite{puska:PRB:1998,lento:JPCM:2003,probert:PRB:2003,corsetti:PRB:2011}
We calculate the formation energy of \{$V$\} from bulk silicon to be
3.43\,eV, which is in excellent agreement with previous calculations
\cite{probert:PRB:2003,corsetti:PRB:2011}.
It is energetically favorable for lithium in the form of \{Li\} and
lithium metal to bind strongly to silicon vacancies already present in
the crystal.
As shown on the right-hand scale of Fig.\,\ref{Fig:Energy}, the
\{4Li,$V$\} complex in scenario (III) has a formation energy of
-0.84\,eV per Li atom, and -1.15\,eV per Li atom when formed from the
\{$V$\} and $T_d$ symmetry interstitial \{Li\} complex (calculated as
the difference between the formation energy of \{Li\} in bulk silicon
and \{4Li,$V$\} in silicon with a \{$V$\} already present).

We did not find any Li/H/V defects that are more stable than
single-impurity--vacancy defects, implying that lithium does not bind
to hydrogen when vacancies are present.

A full study with disordered silicon is beyond the scope of this
article but we have studied the \{4Li,$V$\} complex in two other
tetrahedrally-bonded phases of silicon to assess its stability in
other silicon environments like those found after the first charge-discharge
cycle of a LIB silicon anode.
The BC8 phase \cite{wentorf:science:1963} (with $Ia\overline{3}$
symmetry) has a density of 2.52\,g.cm$^{-3}$, which is higher than
that of diamond-structure silicon (2.3.2\,g.cm$^{-3}$), and we find that the formation
energy in the type (I) scenario of the \{4Li,$V$\} defect is -0.02\,eV
per Li atom, showing that lithium will ``dig-out'' silicon vacancies
in BC8 silicon.
The hypothetical low-energy $P6_122$ framework structure
\cite{pickard:PRB:2010} has a lower density of 2.10\,g.cm$^{-3}$, and
the \{4Li,$V$\} defect has a formation energy -0.05\,eV per Li atom in
the type (I) scenario.
Hence \{4Li,$V$\} is stable over a range of silicon environments with
a range of silicon densities and is therefore likely to be stable in
disordered silicon phases.

Our calculations show that the \{4Li,$V$\} defect is also stable in
diamond-structure ($Fd\overline{3}m$) germanium, with a formation
energy of -0.2\,eV per Li atom, however, due to the smaller lattice
constant of diamond, the defect has a much larger formation energy of
$\sim4$\,eV per Li atom, and it causes considerable stress within the
diamond lattice.

We have presented the structure and electronic properties of the
\{4Li,$V$\} complex.
The  complex is stable once it has formed due to its very
low formation energy, and some experimental evidence exists for such a
defect. \cite{Canham:Physica:1983}
We attribute the stability of \{4Li,$V$\} to it satisfying the Zintl
rules for intermetallic chemical bonding, and other defects such as
\{Li,$V$\}, \{2Li,$V$\} and \{3Li,$V$\} are substantially higher in
energy.
In scenarios I and II, the vacancy is formed alongside the inclusion of lithium. 
This is unlikely to be relevant to the electrochemical lithiation of crystalline silicon.
Scenario III is more physically pertinent since monovacancies already
exist in silicon and \{4Li,$V$\} can form during lithiation.

The \{4Li,$V$\} defect is stable in a range of tetrahedrally-bonded
silicon environments, which suggests that it may also be stable in
disordered phases of silicon.
Scenarios I and II show that \{4Li,$V$\} will form on delithiation of
$a$-Li$_y$Si, where amorphous silicon is formed from high
lithium-content lithium silicides which will have a chemical potential
similar to bulk lithium.
In this case as $a$-Li$_y$Si is delithiated, the silicon recombines to
form $a$-Si, due to its high stability the \{4Li,$V$\} defect is able
to form, trapping both lithium and silicon vacancies in the amorphous
silicon.
The formation of very stable lithium defects may contribute to
incomplete delithiation and a subsequent loss of capacity of a silicon
anode.
Of course, many other lithium-silicon structures will form during the
lithiation of a silicon anode in an LIB, but the \{4Li,$V$\} defect in
silicon is an example of a very stable lithium-silicon structure.

\begin{acknowledgments}
  This work was supported by the Engineering and Physical Sciences
  Research Council (EPSRC) of the U.K.\ E.S.\ acknowledges support
  from a Marie Curie Intra-European Fellowship within the 7th European
  Community Framework Program and thanks Churchill College (Cambridge,
  UK) for a Sackler Research Fellowship. Computational resources were
  provided by the University College London Research Computing
  service.
\end{acknowledgments}

\bibliographystyle{h-physrev}

\begin{thebibliography}{10}




\bibitem{dannefaer:JAP:1995} 
S. Dannefaer and T. Bretagnon,
\newblock J. Appl. Phys. {\bf 77}, 5584 (1995). 

\bibitem{Goldstein:PRL:1966}
B. Goldstein,
\newblock Phys. Rev. Lett. {\bf 17}, 428 (1966).

\bibitem{Klimkova:PSSa:1970}
O.~A. Klimkova and O.~R. Niyazova,
\newblock Phys. Stat. Sol. (a) {\bf 3}, K93 (1970).

\bibitem{wan:JPCM:2010}
W. Wan, Q. Zhang, Y. Cui, and E. Wang,
\newblock J. Phys.: Condens. Matter {\bf 22}, 415501 (2010). 



\bibitem{lai:JES:1976}
S. Lai,
\newblock J. Electrochem. Soc. {\bf 123}, 1196 (1976).

\bibitem{wen:JSSC:1981}
C.~J. Wen and R.~A. Huggins,
\newblock J. Sol. Stat. Chem. {\bf 37}, 271 (1981).

\bibitem{weydanz:JPS:1999}
W.~J. Weydanz, M. Wohlfahrt-Mehrens, and R.~A. Huggins,
\newblock J. Power Sources {\bf 81-82}, 237 (1999).

\bibitem{peng:JPC:2010}
B. Peng, F. Cheng, Z. Tao, and J. Chen,
\newblock J. Chem. Phys. {\bf 133}, 034701 (2010).



\bibitem{kim:JPCC:2010}
H.~Kim, K.~E.~ Kweon, C.-Y. Chou, J.~G. Ekerdt, and G.~S. Hwang,
J. Phys. Chem. C {\bf 114}, 17942 (2010).

\bibitem{watkins:PRB:1970}
G.~D. Watkins and F.~S. Ham,
\newblock Phys. Rev. B {\bf 1}, 4071 (1970).


\bibitem{key:JACS:2009}
B. Key, R. Bhattacharyya, M. Morcrette, V. Sezn\'{e}c, J.-M. Tarascon, and C. P. Grey,
\newblock J. Am. Chem. Soc. {\bf 131}, 9239, (2009).

\bibitem{li:JES:2007}
J. Li and J.~R. Dahn,
J. Electrochem. Soc. {\bf 154}, A156 (2007).





\bibitem{chan:JACS:2012}
M.~K.~Y. Chan, C. Wolverton, and J.~P. Greeley,
\newblock J. Am. Chem. Soc. {\bf 134} 14362  (2012).

\bibitem{goldman:AFM:2011}
J.~L. Goldman, B.~R. Long, A.~A. Gerwith, and R.~G. Nuzzo,
\newblock Adv. Funct. Mater. {\bf 21}, 2412 (2011).


\bibitem{morris:PRB:2011}
A.~J.~Morris, C.~P. Grey, R.~J. Needs, and C.~J. Pickard,
\newblock Phys. Rev. B {\bf 84}, 224106 (2011).

\bibitem{Pickard_AIRSS} 
C.~J. Pickard and R.~J. Needs,
\newblock Phys. Rev. Lett. {\bf 97}, 045504 (2006).
C. J. Pickard and R. J. Needs,
J. Phys.: Condens. Matter \textbf{23}, 053201 (2011). 

\bibitem{morris:PRB:2008}
A.~J.~Morris, C.~J. Pickard, and R.~J. Needs,
\newblock Phys. Rev. B {\bf 78}, 184102 (2008).

\bibitem{morris:PRB:2009}
A.~J.~Morris, C.~J. Pickard, and R.~J. Needs,
\newblock Phys. Rev. B {\bf 80}, 144112 (2009).

\bibitem{mulroue:PRB:2011}
J. Mulroue, A.~J. Morris, and D.~M. Duffy,
\newblock Phys. Rev. B {\bf 84}, 094118 (2011).

\bibitem{mulroue:PRB:2011:He}
J. Mulroue,  M. Watkins, A.~J. Morris and D. Duffy,
\newblock J. Nucl. Mat. {\bf 437}, 1-3, 261 (2012).

\bibitem{CASTEP:ZK:2004}
S.~J. Clark {\em et~al.},
\newblock Z. Kristallogr. {\bf 220}, 567 (2005).

\bibitem{Canham:Physica:1983}
L. Canham, G. Davies, E.~C. Lightowlers, and G.~W. Blackmore,
\newblock Physica {\bf 117B} and {\bf 118B}, 119 (1983).

\bibitem{pickard:PRB:2001}
 C.~J. Pickard and F. Mauri,
\newblock Phys. Rev. B {\bf 63}, 245101 (2001).

\bibitem{yates:PRB:2007}
 J.~R. Yates, C.~J. Pickard, and F. Mauri,      
\newblock Phys. Rev. B {\bf 76} 024401 (2007).  

\bibitem{ivanov:ACIE:2012}
A.~S. Ivanov, A.~J. Morris, K.~V. Bozhenko, C.~J. Pickard, and A.~I. Boldyrev,
\newblock Angew. Chem. Int. Ed. {\bf 51}, 8330 (2012).

\bibitem{optados}
A.~J. Morris, R.~J. Nicholls, C.~J. Pickard, and J.~R. Yates 
\newblock \textsc{Optados} User’s Guide, version 1.0.0 (University of Cambridge,
Cambridge, UK, 2013).
 R. J. Nicholls, A. J. Morris, C. J. Pickard, and J. R. Yates, 
\newblock J. Phys.: Conf. Ser. {\bf 371} 012062 (2012).

\bibitem{puska:PRB:1998}
M.~J. Puska, S. P\"{o}ykk\"{o}, M. Pesola, and R.~M. Nieminen,
\newblock Phys. Rev. B {\bf 58}, 1318 (1998).

\bibitem{lento:JPCM:2003}
J. Lento and R.~M. Nieminen,
\newblock  J. Phys.: Condens. Matter {\bf 15}, 4387 (2003). 

\bibitem{probert:PRB:2003}
M.~I.~J. Probert and M.~C. Payne,
\newblock Phys. Rev. B {\bf 67}, 075204 (2003).

\bibitem{corsetti:PRB:2011}
F. Corsetti and A.~A. Mostofi,
\newblock Phys. Rev. B {\bf 84}, 035209 (2011).

\bibitem{wentorf:science:1963}
R.~H. Wentorf and J.~S. Kasper, 
\newblock Science {\bf 139}, 338 (1963).

\bibitem{pickard:PRB:2010}
C.~J. Pickard and R.~J. Needs,
\newblock Phys. Rev. B {\bf 81}, 014106 (2010).



\end{thebibliography}

\end{document}